\documentstyle[aps,psfig,manuscript]{revtex}
\draft
\begin{document}
\title{Interaction of a granular stream with an obstacle}
\author{Volkhard Buchholtz and Thorsten P\"oschel\footnote{volkhard@summa.physik.hu-berlin.de, thorsten@itp02.physik.hu-berlin.de}} 
\address{Humboldt-Universit\"at zu Berlin, Institut f\"ur Physik,\\ Invalidenstra\ss e 110, D--10115 Berlin, Germany}
\date{\today}
\maketitle
\begin{abstract}
We investigate numerically the interaction of a stream of granular particles with a resting obstacle in two dimensions. For the case of high stream velocity we find that the force acting on the obstacle is proportional to the square of the stream velocity, the density and the obstacle size. This behavior is equivalent to that of non-interacting hard spheres. For low stream velocity there appears a gap in between the obstacle and the incoming stream particles which is filled with granular gas of high temperature and low density. As soon as the gap appears the force does not depend on the square of velocity of the stream but the dependency obeys another law.
\end{abstract}
\section{Introduction}
When a stream of solid frictionless particles $i$ with radii $R_i\rightarrow 0$ which do not interact with each other is scattered at an obstacle one expects from momentum conservation~\cite{Boltzmann} that the net force $F$ acting on the obstacle obeys the law
\begin{equation}
F\sim \rho \cdot A \cdot v_{st}^2~. 
\label{fboltzmann}
\end{equation}
 $A$ is the projection of the obstacle to the stream direction, $v_{st}$ is the velocity of the stream particles, and $\rho$ is the density of the stream. When the stream consists of granular particles which {\em do} interact with each other, i.e. between which forces and torques are transferred, and which undergo damping during collisions one can expect deviations from the behavior in eq.~\ref{fboltzmann}. The aim of this paper is to investigate the interaction between the stream and the obstacle using soft particle molecular dynamics in two dimensions. As far as we know the described phenomena have not been investigated experimentally so far.

The paper is organized in three parts. In section \ref{sec:exp} we explain the numerical experiment in detail. Section~\ref{sec:scatter} contains our results concerning the angular distributions of the scattered stream, i.e. the distributions of mass, momentum and angular momentum as functions of the scatter angle $\Theta$. In \ref{sectioninteraction} we investigate the force experienced by the obstacle from the impacting granular stream, whereas in section \ref{sec:fields} we present some continuum properties of the experiment, i.e. the fields of granular temperature and density. In the final section \ref{sec:discussion} we discuss the results.

\section{The numerical experiment} 
\label{sec:exp}
\begin{figure}[htb]
\centerline{\psfig{figure=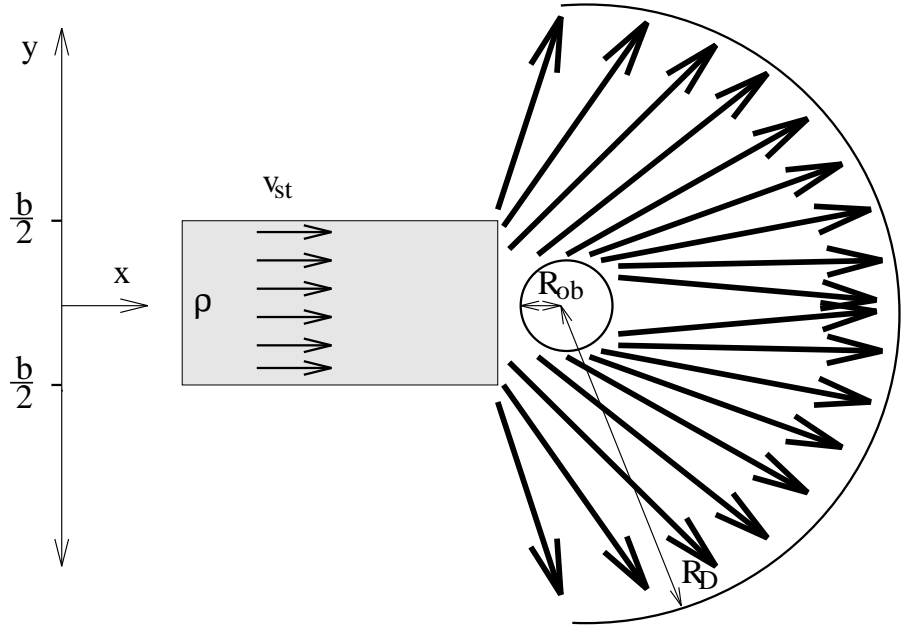,width=12cm,angle=270}}
\caption{The setup of the numerical experiment.}
\label{setup}
\end{figure}

The setup of the numerical experiment is sketched in fig.~\ref{setup}, $b$ is the width of the stream, $v_{st}$ the velocity of the stream particles. When the obstacle is a sphere we can describe it by its radius $R_{ob}$. At distance $R_D$ from the center of the obstacle with $R_{ob}\ll R_D$ there is a screen where we measure the angle dependent distributions of the scattered stream. In the simulation we used $R_D \ge 12\cdot R_{ob}$. The radii of the stream particles $R_i$ are equally distributed in the interval $[0.05\,;\,0.11]\,\mbox{cm}$. Each of them has translational and rotational degrees of freedom.

Before interacting with the obstacle the particles move in an uniform stream (left in fig.~\ref{setup}), all  with the same initial velocity $v_{st}$. When released the particles move towards the obstacle, collide with the obstacle and once cross the screen plane (right in fig.~\ref{setup}). After crossing the screen plane the particles are removed from the calculation.

Initially the stream particles are arranged in columns. All particles which belong to the same column have the same $x$-position. One particle of each column has been set to a random $y$-position, then the column was filled in positive and negative $y$-direction up to $y_{\mbox{\it\footnotesize\it min}}=-b/2$ and $y_{\mbox{\it\footnotesize\it max}}=b/2$ respectively. At the initialization time the distances between the centers of two neighboring spheres with radii $R_i$ and $R_j$ in horizontal and vertical direction $\Delta x$ and $\Delta y$ are
\begin{mathletters}
\begin{eqnarray}
\Delta x &=& 2\cdot R^{\mbox{\it\footnotesize\it max}} \sqrt{1/\rho} \\
\Delta y &=& \left(2\cdot R^{\mbox{\it\footnotesize\it max}} + 0.2*\mbox{GRND} \right)\sqrt{1/\rho}~,
\end{eqnarray}
\label{delta}
\end{mathletters}
where $\rho$ is the density of the stream as used in eq.~\ref{fboltzmann}, $R^{\mbox{\it\footnotesize\it max}}$ is the radius of the largest stream particle and $\mbox{GRND}$ is a Gaussian random number with standard deviation of one. All particles move with the same initial stream velocity $v_{st}$ in positive $x$-direction. The density value $\rho=1$ in eqs.~\ref{delta} means that the distance between neighboring columns is $2~R^{\mbox{\it\footnotesize\it max}}$. According to the described method the number of particles is not conserved during the simulation. Particles are removed from the calculation when they cross the screen plane, and particles are added to the system according to the initialization scheme. The total number of particles which are involved in the simulation varies between $N=4000$ and $N=6000$.

Two particles $i$ and $j$ at positions $\vec{r}_i$ and $\vec{r}_j$ interact if the distance between their center points is smaller than the sum of their radii. In this case we define the particle overlap $\xi_{ij}= R_i+R_j-\left|\vec{r}_i - \vec{r}_j\right|$. Two colliding spheres feel the force
\begin{equation}
\vec{F}_{ij}= F_{ij}^{N}\cdot \vec{n}^N + F_{ij}^{S}\cdot \vec{n}^S
\label{kraft}
\end{equation}
with the unit vectors in normal direction $\vec{n}^N=\frac{\vec{r}_i -\vec{r}_j}{\left| \vec{r}_i - \vec{r}_j\right|}$ and in shear direction $\vec{n}^S$, $\left| \vec{n}^S \right|=1$, $\vec{n}^N \cdot \vec{n}^S = 0$. 

In viscoelastic approximation the normal force acting between colliding spheres has been derived rigorously by Brilliantov et al.~\cite{BSHP}
\begin{equation}
 F_{ij}^N = \frac{Y\sqrt{R^{\,\mbox{\it\footnotesize\it eff}}_{ij}}}{\left( 1-\nu ^2\right)} ~\left(\frac{2}{3}\xi^{3/2} + A \sqrt{\xi}\, \dot {\xi} \right)
\label{normal}
\end{equation}
with
\begin{equation}
A =  \frac 13 \frac{\left(3\eta_2-\eta_1\right)^2} {\left( 3\eta_2+2\eta_1\right)} \frac{\left(1-\nu ^2\right)\left(1-2~\nu \right)} {Y\,\nu^2}~.
\end{equation}
 $Y$, $\nu$ and $\eta_{1/2}$ are the Young-Modulus, the Poisson ratio and the viscous constants of the particle material. $R^{\,\mbox{\it\footnotesize\it eff}}_{ij}$ is the effective radius of the spheres $i$ and $j$
 \begin{equation}
R_{ij}^{\mbox{\it\footnotesize\it eff}}= \frac{R_i \cdot R_j}{R_i + R_j}\,.
 \end{equation}

The functional form of eq.~(\ref{normal}) has been found before by Kuwabara and Kono~\cite{KuwabaraKono} using heuristic arguments and has been derived recently by Morgado and Oppenheim~\cite{MorgadoOppenheim:1997} from first principle calculations.

For the shear force we apply the Ansatz by Haff and Werner~\cite{HaffWerner} which has been proven to give correct results in many other molecular dynamics simulations of granular material, e.g.~\cite{HW1,HW2,HW3,HW4}
\begin{equation}
F_{ij}^S = \mbox{sign}\left({v}_{ij}^{\,\mbox{\it\footnotesize\it rel}}\right) \min \left\{\gamma_S \cdot m_{ij}^{\,\mbox{\it\footnotesize\it eff}} \cdot \left|{v}_{ij}^{\,\mbox{\it\footnotesize\it rel}}\right|~,~\mu \cdot \left|F_{ij}^N\right| \right\} ~.
\label{shear}      
\end{equation}
The relative velocity ${v}_{ij}^{\,\mbox{\it\footnotesize\it rel}}$ between the particles $i$ and $j$ with angular velocities $\Omega_i$ and $\Omega_j$ and the effective mass $m_{ij}^{\,\mbox{\it\footnotesize\it eff}}$ of the particles $i$ and $j$ are given by
\begin{eqnarray}
{v}_{ij}^{\,\mbox{\it\footnotesize\it rel}} &=& \left(\dot{\vec{r}}_i - \dot{\vec{r}}_j\right)\cdot \vec{n}^S + R_i \cdot {\Omega}_i + R_j \cdot {\Omega}_j ~,\\
m_{ij}^{\,\mbox{\it\footnotesize\it eff}} &=& \frac{m_i \cdot m_j}{m_i + m_j} ~.
\label{eq_eff_mass}
\end{eqnarray}
The resulting torques $M_i$ and $M_j$ acting upon the  particles are
\begin{equation}
M_i = F_{ij}^S \cdot R_i~~,~~ M_j = - F_{ij}^S \cdot R_j ~.
\end{equation}
Eq.~(\ref{shear}) takes into account that the particles slide upon each other for the case that the inequality 
\begin{equation}
\mu \cdot \mid F_{ij}^N \mid~<~\mid F_{ij}^S \mid
\end{equation}
holds, otherwise they feel some viscous friction.

The particles interact with the spherical obstacle in the same manner as the particles with each other, i.e. the force $\vec{F}_{ij}$ acting on the $i$th particle when colliding with the obstacle is given by eq.~(\ref{kraft}), with $R_j=R_{ob}$ and $\vec{r}_j$ is the position of the obstacle. The mass of the obstacle is infinite, hence the effective mass (eq.~(\ref{eq_eff_mass})) is $m^{\mbox{\it\footnotesize\it eff}}=m_i$. The flat obstacle was built up of equal spheres as shown below in fig.~\ref{snapflat} with the same material constants as the stream particles but of infinite mass.  Their radii have been $R_{ob}^*=\overline{R_i}$, where $\overline{R_i}$ is the mean radius of the stream particles. The force acting on a stream particle when colliding with such a particle is calculated according to eq.~(\ref{kraft}).

For the integration of Newton's equation of motion we used the classical Gear predictor-corrector molecular dynamics algorithm \cite{Gear,Allen} of 6th order for the positions and of 4th order for the rotation of the particles.

In the numerical simulation we used the parameters given in Table~\ref{parameter} which are valid for a typical granular material. The time step for the numerical integration of the Gear predictor-corrector scheme was $\Delta t=1\cdot10^{-5}\,\mbox{sec}$. Using this time step the system behaves numerically stable, i.e. tests with time step $\Delta t=2\cdot10^{-5}\,\mbox{sec}$ lead to the same results.

The simulations presented in this paper have been performed on the 16 processor message passing parallel computer {\it KATJA} (http://summa.physik.hu-berlin.de/KATJA/).

\section{Results}
\subsection{Scattering of the granular stream}
\label{sec:scatter}
First we investigate the angle dependent distributions of the scattered particles for a numerical experiment with a fixed spherical obstacle  of radius $R_{ob}= 1.5\,\mbox{cm}$. The differential mass intensity $dm$ of the scattered stream normalized by the incoming stream is defined by 
\begin{equation}
dm(\Theta) = \frac{dm_D(\Theta)}{m_{st}} ~,
\label{diffintensity}
\end{equation} 
where $dm_D(\Theta)$ is the mass of the particles which are scattered into the interval $[\Theta, \Theta+d\Theta ]$ and $m_{st}$ is the total mass of the incoming particles. $\Theta=0$ is the direction directly behind the obstacle.

\begin{figure}[htb]
\centerline{\psfig{figure=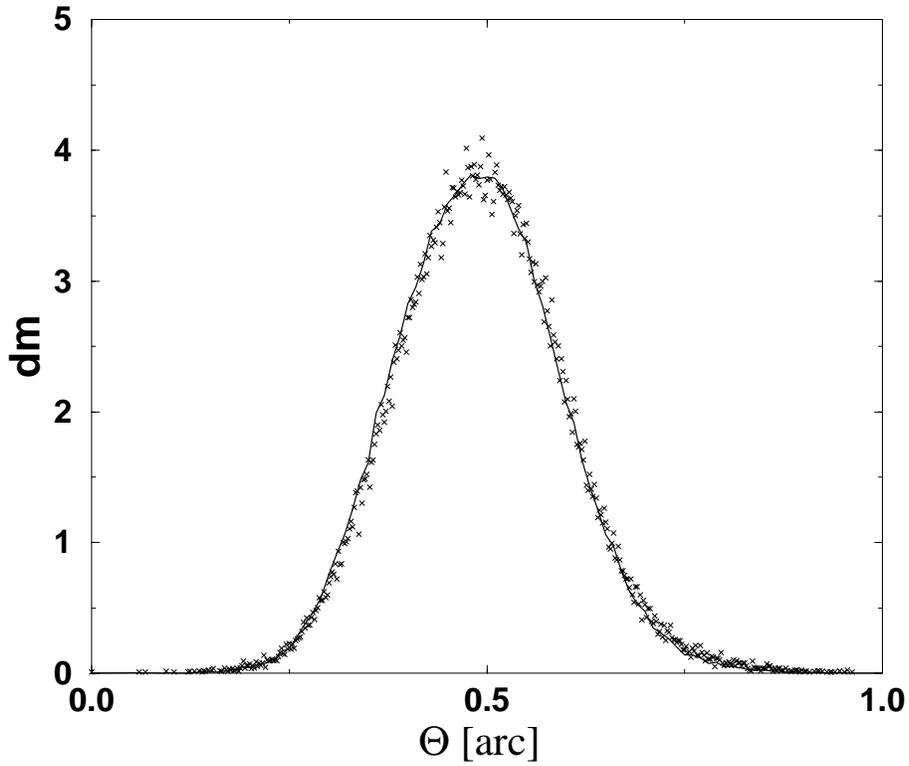,width=12cm,angle=270}} 
\caption{The differential intensity of the stream scattered at a spherical obstacle over the angle $\Theta$.}  
\label{intensity}
\end{figure}

Figure~\ref{intensity} shows the differential intensity, i.e.~the mass $dm(\Theta)$ of the particles which are scattered into a certain interval $[\Theta, \Theta+\Delta\Theta ]$,  as a function of the scatter angle $\Theta$. The problem is symmetric due to the symmetry of the incoming particle stream, hence the intensity is symmetric as well. For the sake of better statistics we plot in fig.~\ref{intensity} $dm(\Theta) + dm(-\Theta)$ over $\Theta$. We find maxima at $\Theta \approx \pm 0.5 = 29 ^o$. The minimum at $\Theta \approx 0$ is due to the finite size of the obstacle, so only few particles are scattered into the region directly behind the obstacle. 

\begin{figure}[htb]
\centerline{\psfig{figure=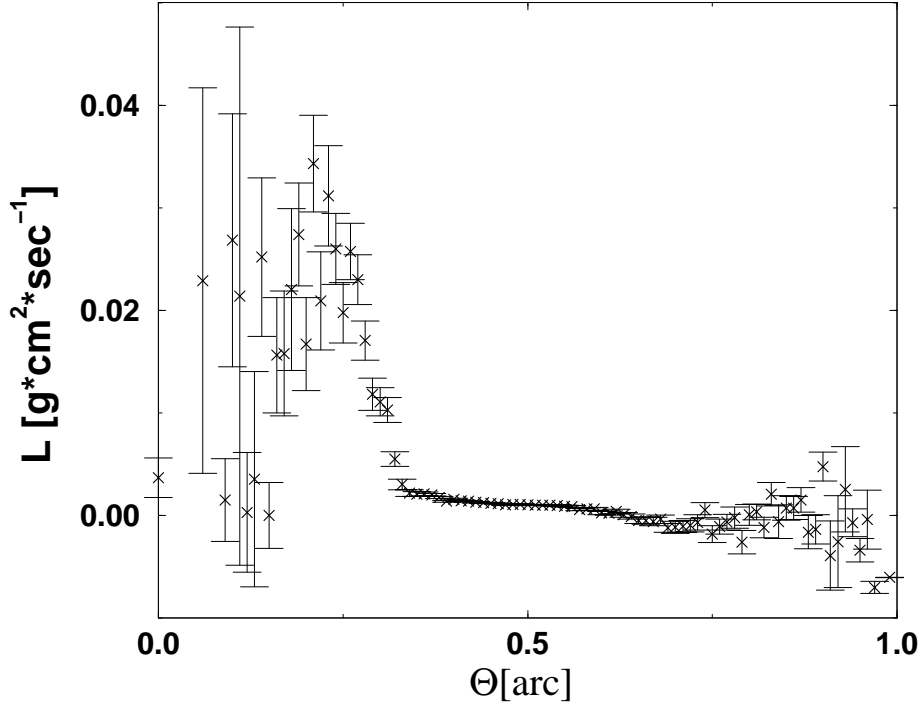,width=12cm,angle=270}} 
\caption{The distribution of angular momentum $L(\Theta)$ over the scatter angle $\Theta$.} 
\label{rotation}
\end{figure}

Due to the different sense of the transferred angular moments resulting from collisions at the lower or at the upper side (in $y$-direction) of the particles the angular momentum  $L(\Theta)$ of the particles which fall into a certain interval of scatter angle $\Theta+d\Theta$ has a negative sign for $\Theta < 0$ and a positive for $\Theta > 0$. Hence, the distribution is antisymmetrically. As in the case of the intensity we find symmetry of the absolute value of the distribution of the angular velocity. Therefore, we draw in fig.~\ref{rotation} $L(\Theta) - L(-\Theta)$ over $\Theta$. When the particles collide with the obstacle they lose kinetic energy due to damping. The larger the distance $|y|$ of the particles in vertical direction from the obstacle (see fig.~\ref{setup}) the smaller is the dissipated kinetic energy.  Hence, the particles moving with large $y$ (outer particles) are faster than others with lower $|y|$ (inner particles). Therefore, the stream particles undergo more intensive collisions from the outer side (large $|y|$) than from the inner (smaller $|y|$). Particles scattered into the region behind the obstacle have a higher angular velocity since they need to be pushed by faster particles multiple times from the same side to reach this region. These effects are the reason for the distribution $L(\Theta)$ of the angular momentum. In agreement with the above arguments fig.~\ref{rotation} shows that the distribution of the absolute value of the averaged angular momentum $L(\Theta)$ of particles decreases with increasing $\Theta$. In the range $\Theta \approx 0$ the opposite rotational sense for particles from the upper and from the lower side combine to an angular momentum of zero. In this region the error bars are large according to the bad statistics.

\begin{figure}[htb]
\centerline{\psfig{figure=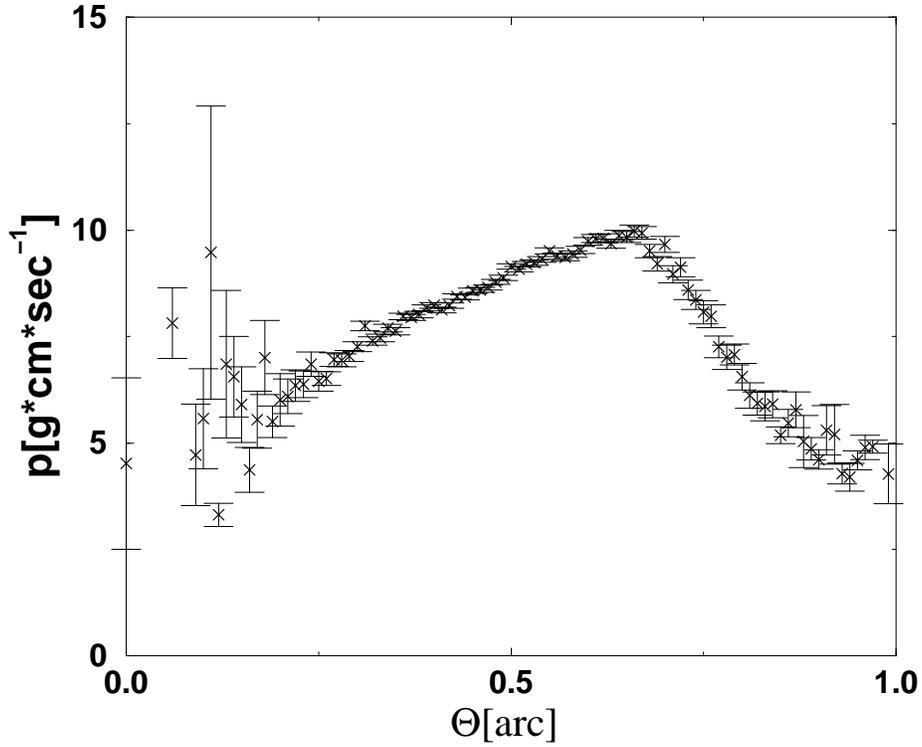,width=12cm,angle=270}} 
\caption{The distribution of the absolute value of the averaged particle momentum over the scatter angle $\Theta$. Again, for a better statistics we took advantage of the symmetry of the problem.}
\label{vdistribution}
\end{figure}

The averaged absolute value of the momentum $\left| m_i \vec{v}_i \right|= m_i \left[ \left( v^{(x)}_i \right)^2 + \left( v^{(y)}_i \right)^2 \right]^{0.5}$ of the scattered particles as function of the scatter angle is drawn in fig.~\ref{vdistribution}. The velocity of the input stream is $v_{st}=500 \,\mbox{cm/sec}$. As described above the energy loss of the stream is higher for particles scattered by low angles $\Theta$. In this region we find large error bars according to bad statistics -- only few particles are scattered directly behind the obstacle. The maxima of the distribution $p(\Theta)$ are close to the intensity maxima. There seems to be a preferred region for the scatter angle where the traces of the particles are almost in parallel, i.e. where energy dissipation is low. We will come back to this presumption when we discuss the granular temperature field of the stream.  

\subsection{Force experienced by the obstacle}
\label{sectioninteraction}
The impact of the stream particles on the target results in an effective force between the stream and the obstacle. With the strong simplifying assumptions that the particles have no rotational degree of freedom and the particles interact only with the obstacle but not with each other one obtains from kinetic gas theory a theoretical approximation for the force $F$  
\begin{equation}
F = \frac{dp}{dt}  \sim R_{ob}\,\rho\,v_{st}^2~. \label{ftheor}
\end{equation}
The force is the transferred momentum per time which is the mass of the particles colliding with the obstacle in the time $dt$ times the stream velocity.  Eq.~(\ref{ftheor}) implies the relations
\begin{mathletters}
\begin{eqnarray}
&F \sim& \rho  \label{Frho}~,\\
&F \sim& R_{ob}  \label{FRadius}~,\\
\mbox{and}~~&F \sim& v_{st}^2 \label{Fvsq}~. 
\end{eqnarray}
\label{F}
\end{mathletters}
In the following we will investigate each of the relations~(\ref{F}) 
numerically for granular particles for which the simplifying assumptions are evidently {\em not} provided.

The force acting on the obstacle fluctuates according to the irregularities of the stream, hence, to get reliable values we averaged the force for each data point over a certain time that corresponds to approximately 500 rows of stream particles colliding with the obstacle. As shown in the results this time is sufficient to generate very small error bars. 

\begin{figure}[htb]
\centerline{\psfig{figure=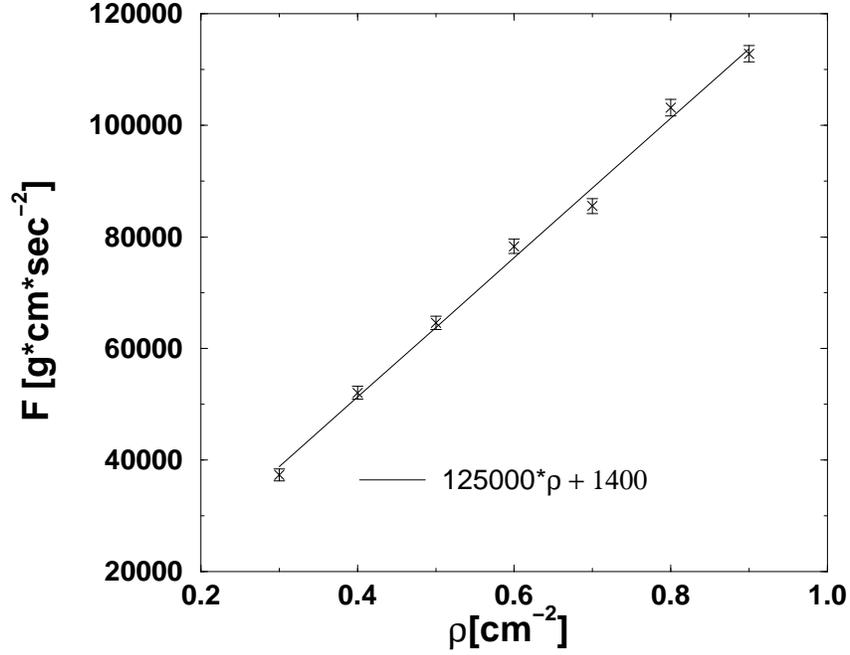,width=11cm,angle=270}} 
\caption{The force over the stream density $\rho$.}
\label{foverdichte}
\end{figure}

As we see from fig.~\ref{foverdichte} the simulation is in good agreement with eq.~(\ref{Frho}) for the force $F$ as a function of the stream density $\rho$ of the input stream. The function $F=F(\rho)$ can be fitted to
\begin{equation}
F=(125000 \cdot \rho+1400)\,\frac{\mbox{g}\,\mbox{cm}}{\mbox{sec}^2}~.
\label{FRho}
\end{equation}

\begin{figure}[htb]
\centerline{\psfig{figure=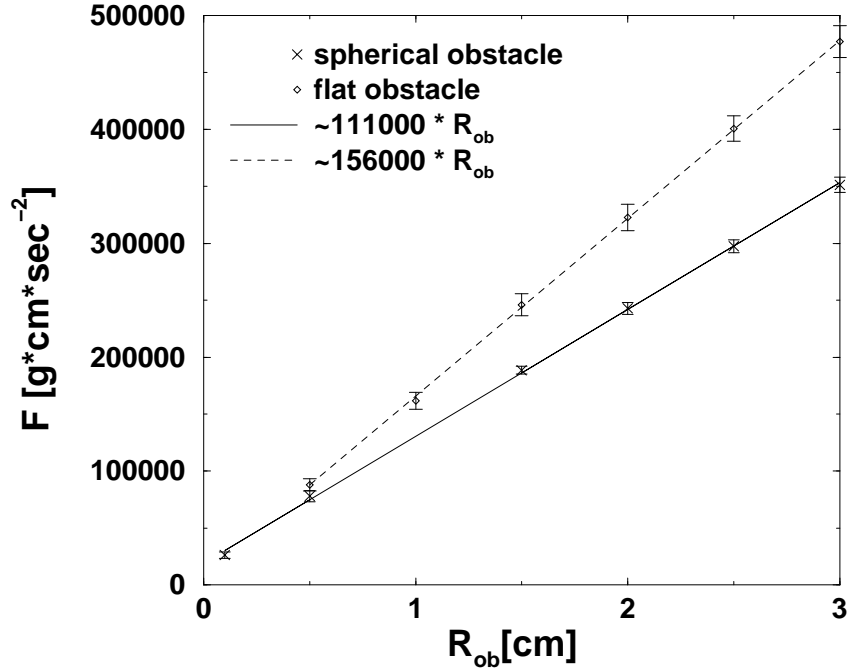,width=11cm,angle=270}} 
\caption{The force over the obstacle size $R_{ob}$. In agreement with eq.~(\ref{FRadius}) we find a linear function. The solid line shows the force experienced by a spherical obstacle, the dashed line shows the force for a flat obstacle (explanation see below).}
\label{foverradius}
\end{figure}

In agreement with eq.~(\ref{FRadius}) the acting force $F$ is also linear in its dependency on the radius $R_{ob}$ of the obstacle as drawn in  fig.~\ref{foverradius} (crosses and solid line). It can be expressed by the fit
\begin{equation}
F=(111 \cdot R_{ob} + 18)\,10^3\,\frac{\mbox{g}\,\mbox{cm}}{\mbox{sec}^2}~.
\label{FR}
\end{equation}

There exists a resulting force $F \approx 18\cdot 10^3\,\mbox{g}\cdot \mbox{cm/sec}^2$ for $R_{ob} \rightarrow 0$ since as shown below even a {\em very} small obstacle hinders the stream particles (of finite radius) and results in a certain resistance for the stream.

Thus, we find the surprising result that the functional form of the acting force as a function of the density of the stream $\rho$ (eq.~(\ref{FRho})) and of the radius $R_{ob}$ of the obstacle (eq.~(\ref{FR})) which we measured in our simulations equals that of the crude approximation eq.~(\ref{ftheor}). This coincidence is not trivial since the stream particles which collide with the obstacle build up a characteristic ``corona'' around the obstacle as it can be seen in fig.~\ref{fig}.  The size and shape of this corona vary with the stream density $\rho$ and the radius $R_{ob}$.

For the case of a flat obstacle one observes qualitatively the same behavior as for a spherical one. The snapshot fig.~\ref{snapflat} shows that in this case the corona which forms the effective obstacle has approximately the same shape as for spherical obstacles. For the force as a function of the obstacle size we measure
 $F_{\mbox{\it\footnotesize\it flat}}\sim 156 \cdot 10^3 \cdot R_{ob}$ instead of $F \sim 111 \cdot 10^3\cdot R_{ob}$ for the simulation with a spherical obstacle of the same size and identical simulation parameters (fig.~\ref{foverradius}, dashed line). The shape of the obstacle causes only a change in the prefactor of the force law but leaves the functional form unaffected.

The color in figs.~\ref{fig} and \ref{snapflat} codes for the absolute value of the particle's velocity. In both cases there is a wide region in front of the obstacle where the velocities are very small. In this region the particles almost agglomerate to the obstacle. The width of this region depends strongly on the stream velocity and on the material parameters. In figure~\ref{snapflat} one observes a small gap of low particle density directly in front of the obstacle. In this gap one finds particles of high kinetic energy. The properties of the gap will be discussed below.

\begin{figure}[htb]
\centerline{\psfig{figure=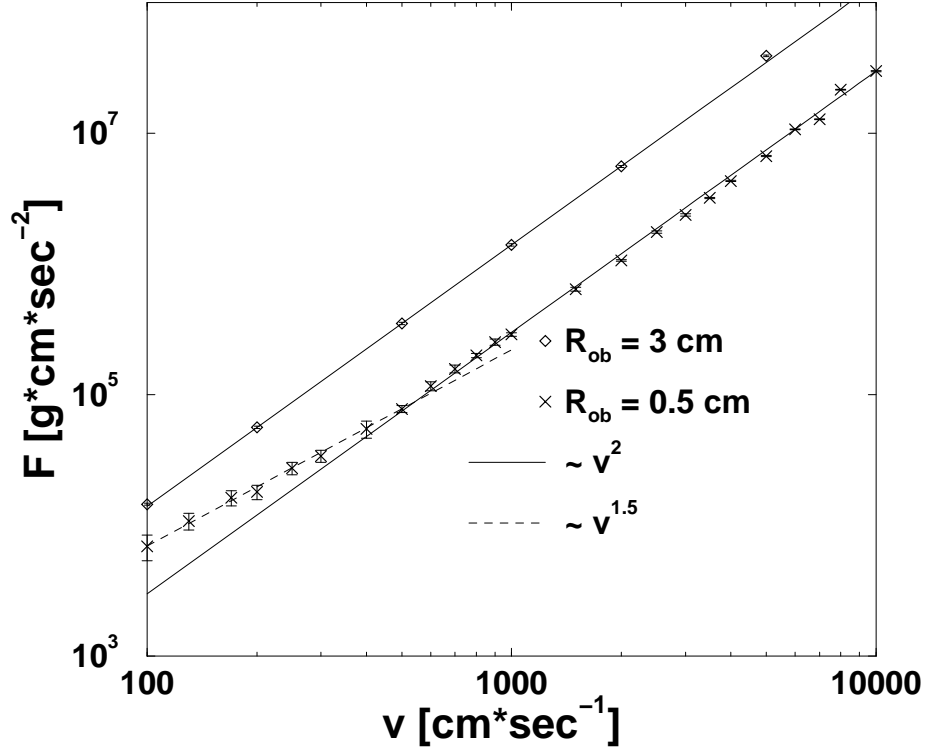,width=12cm,angle=270}} 
\caption{The force over the stream velocity $v_{st}$. Small obstacle: for large velocity $v_{st}$ the force agrees with the crude approximation (\ref{Fvsq}). For smaller impact velocity, however, the force deviates from this function. The dashed line shows the function $F\sim v^{3/2}$. Large obstacle: the quadratic function remains valid even for smaller $v_{st}$. We expect a deviation from this function for yet smaller impact velocity.}
\label{foverv}
\end{figure}

The dependence of the force $F$ on the velocity $v_{st}$ of the stream is more complicate.  Fig.~\ref{foverv} shows that the quadratic behavior from the estimation eq.~(\ref{Fvsq}) is only valid for high velocities ($v_{st} > 700 \, \mbox{cm/sec}$) in the case of a small obstacle. For lower velocities the curve deviates significantly from this function. The data can be fitted to
\begin{equation}
F\sim v^{1.5}~,
\label{FV}
\end{equation}
however, we are not sure whether a power law is the adequate representation of these data.

Hence,  from eqs.~(\ref{FRho}), (\ref{FR}) and (\ref{FV}) we find for our particular system 
\begin{equation}
F \sim \rho \cdot R_{ob} \cdot \left\{\begin{array}{ll} 
v_{st}^2 &  v_{st} > 700\, \mbox{cm/sec} \\
v_{st}^{1.5} & v_{st} < 700\, \mbox{cm/sec}~.\end{array} \right.
\end{equation}

For large impact velocity $v_{st}$ the corona is in direct contact with the obstacle as shown in the snapshots (figs.~\ref{fig} and \ref{snapflat}). In the region of the transition where the force starts to deviate from the quadratic law we observe that the agglomerated region of particles having low velocity in front of the obstacle becomes large. In figs.~\ref{fig} and \ref{snapflat} the size of this region is determined mainly by the size of the obstacle, whereas for small stream velocity the ``cold'' region may become significantly larger than the obstacle itself. Fig.~\ref{fig:gapsnaps} shows three snapshots of the system with stream velocity well above (left), at the transition point (middle) and well below the transition (right). 

Moreover close to the transition one observes a gap forming in between the corona and the obstacle. This gap is filled with granular gas at high temperature. In section \ref{sec:fields} we will discuss the field of granular temperature which supports this statement.

\begin{figure}[htb]
\centerline{\psfig{figure=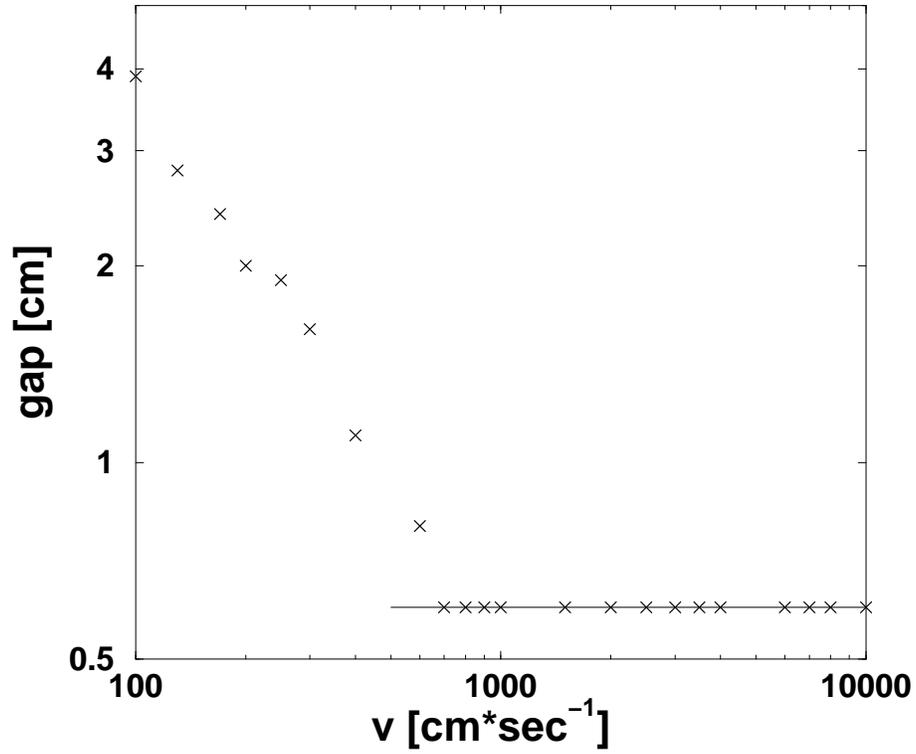,width=12cm,angle=270}} 
\caption{The size of the gap as a function of the stream velocity. The value of $v_{st}$ when the gap appears coincides with the stream velocity when the force which acts on the obstacle deviates from the $F\sim v_{st}^2$-behavior.}
\label{luecke}
\end{figure}

The width of this gap depends on the stream velocity as shown in fig.~\ref{luecke}. For $v_{st} \gtrsim 700 \,\mbox{cm/sec}$ there is no gap. It appears at $v_{st} \approx 700 \,\mbox{cm/sec}$ and there is a sharp transition. The width of the gap increases for decreasing stream velocity. A similar effect, i.e. the appearance of a gap when varying the driving parameters of the system,  was found recently in an event driven simulation of granular particles which are heated (with granular temperature) from below when gravity acts~\cite{EPR}. Similar as in our case there seems to be a critical value for the driving parameter (in~\cite{EPR} the temperature) when the gap appears.
 
For the region of parameters applied in our simulations the stream width $b=16 \mbox{cm}$ does not influence the results. For obstacle size $R_{ob}=0.5\, \mbox{cm}$, stream velocity $v_{st} = 200\, \mbox{cm/sec}$ 
and stream density $\rho = 1$
we made tests with $b_1=2~b$. We found for the time averaged force acting on the obstacle $\bar{F}(b_1)=1.009~\bar{F}(b)$ which fits well into the region covered by the error bars.

\section{Continuum representation of the scatter experiment}
\label{sec:fields}

It has been shown by several authors (e.g.~\cite{Haff:1983,GoldhirschZanetti:1993,GoldhirschSela,EsipovPoeschel:1995,LunSavage:1987,SEP,SPE}) that under certain conditions the dynamics of granular material can be described using hydrodynamics equations. In this section we want to present field data for the density $\rho\left(\vec{r}\right)$ and the temperature $T\left(\vec{r}\right)$ of the granular system. The field data have been generated by coarse graining the particle data to a grid with grid constant $a = 0.1 \mbox{cm} \approx R^{\mbox{\it\footnotesize\it max}}$. The average has been taken over a time interval which corresponds to approximately 500 rows of stream particles colliding with the obstacle.

Fig.~\ref{fig:fields} shows the density and velocity fields as color plots for $v_{st}=500 \mbox{cm/sec}$ and three different radii of the spherical obstacle. Red color stands for high and blue for low values of the density or the velocity. All figures in one row have the same scaling for the colors. For the case of the large sphere ($R_{ob}=1.5 \mbox{cm}$) the density in front of the obstacle is very high. The particles which join this high density region are of low temperature. The temperature rises up to a characteristic distance from the surface of the obstacle and has its maximum in the region where the incoming stream particles collide with the corona which surrounds the obstacle, i.e. the location of maximum temperature coincides with the boundary of the corona. As discussed in the context of figs.~\ref{fig} and \ref{snapflat} for smaller obstacle size (or lower stream velocity) a region of low density and high temperature forms in front of the obstacle, i.e. the highest density can be found in a certain distance from the obstacle. The region of highest temperature does not necessarily coincide with the border of the corona.

The fields of granular temperature and density as well as the appearance of the gap are reminiscent to the behavior of a bullet flying in air with supersonic velocity. It has been shown before that rapid granular flows may contain spatial regions in which the flow is supersonic~\cite{EsipovPoeschel:1995,TanGoldhirsch,EsipovPC,GoldshteinShapiroGutfinger3:1996}. The properties of the flow and the possible embedding into a continuum mechanics theory are subjects of work in progress~\cite{BEP}.

\section{Discussion}
\label{sec:discussion}
We investigated numerically the properties of a stream of granular particles which is scattered by an obstacle and the force acting upon the obstacle as a function of the stream density $\rho$, the stream velocity $v_{st}$ and the obstacle size $R_{ob}$ using molecular dynamics technique. The fields of granular temperature and density have been generated from the molecular dynamics data by coarse graining. 

Surprisingly we find the force to be proportionally to the density $\rho$, the obstacle size $R_{ob}$, and for high stream velocities proportionally to $v_{st}^2, $ i.e. $F\sim \rho\,R_{ob}\,v_{st}^2$ as one would expect from a stream of non-interacting hard spheres. For lower velocities we find that the force deviates from the $F\sim v_{st}^{2}$ law, but still we find a linear law for the force as a function of $R_{ob}$ and $\rho$. This deviation occurs at a critical value for the stream velocity $v_{st}$. At the same velocity we observe a gap forming in between the surface of the obstacle and the corona of stream particles which agglomerate and rest around the obstacle. This gap rises when the stream velocity decreases. The widths of this gap which is filled with granular gas at high temperature grows with decreasing stream velocity.

This agglomeration of particles in front of the obstacle leads to an effective obstacle size $R^{\mbox{\it\footnotesize\it eff}}_{ob}$ larger than $R_{ob}$. The shape of the corona does not sensitively depend on the shape of the obstacle, the functional form of the force $F\sim \rho\,R_{ob}\,v_{st}^2$ remains invariant if the spherical obstacle is replaced by a flat one.

\begin{figure}[htb]
\centerline{\psfig{figure=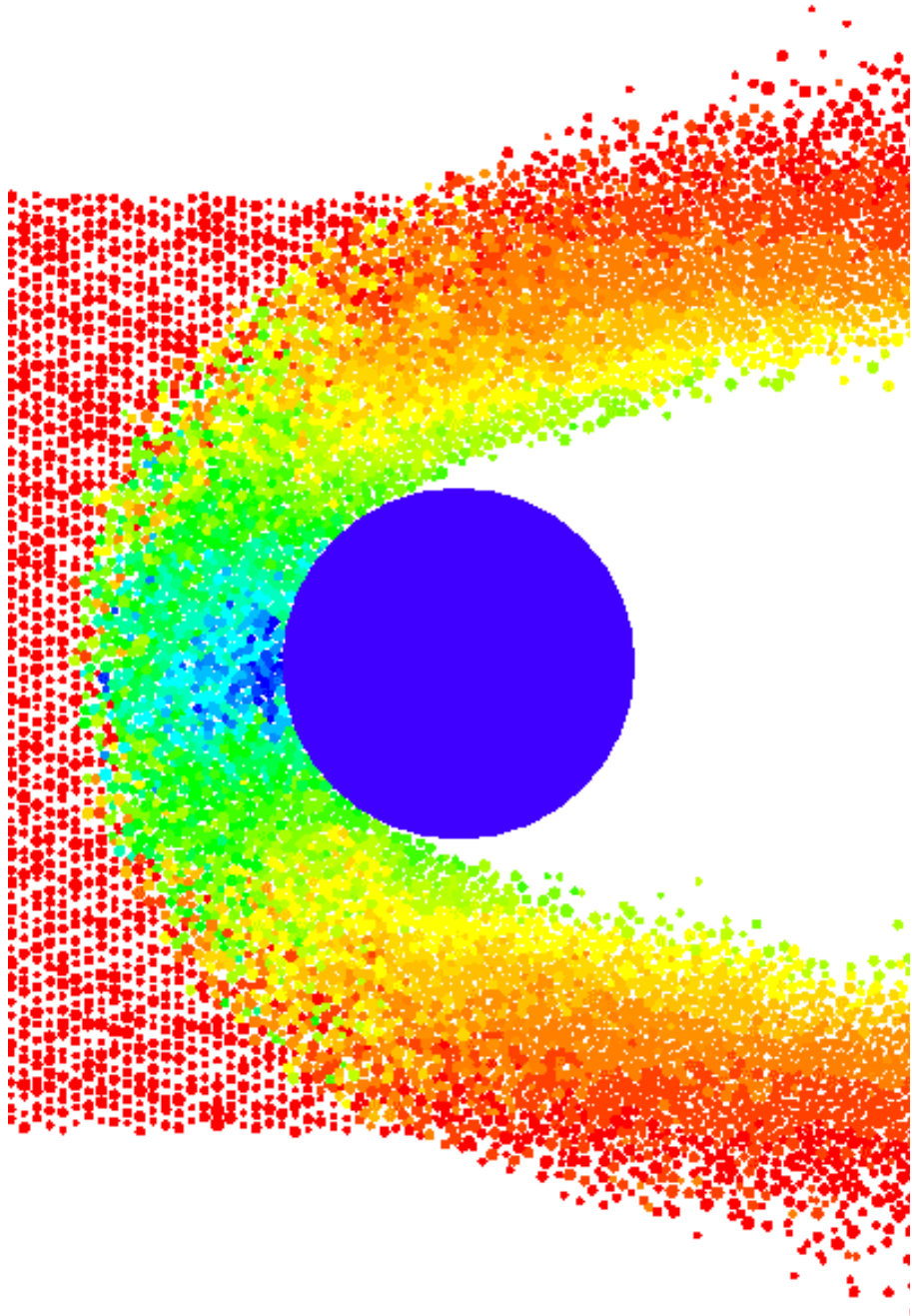,width=12cm,angle=0,clip=}}
\vspace{0.3cm}
\caption{Snapshot of the simulation. The dissipation during the collisions causes a stationary corona which shape and size depend on the parameters of the simulation. The color codes for the absolute value of the particle velocities, red means high velocity, blue means low.}
\label{fig}
\end{figure}

\begin{figure}[htb]
\centerline{\psfig{figure=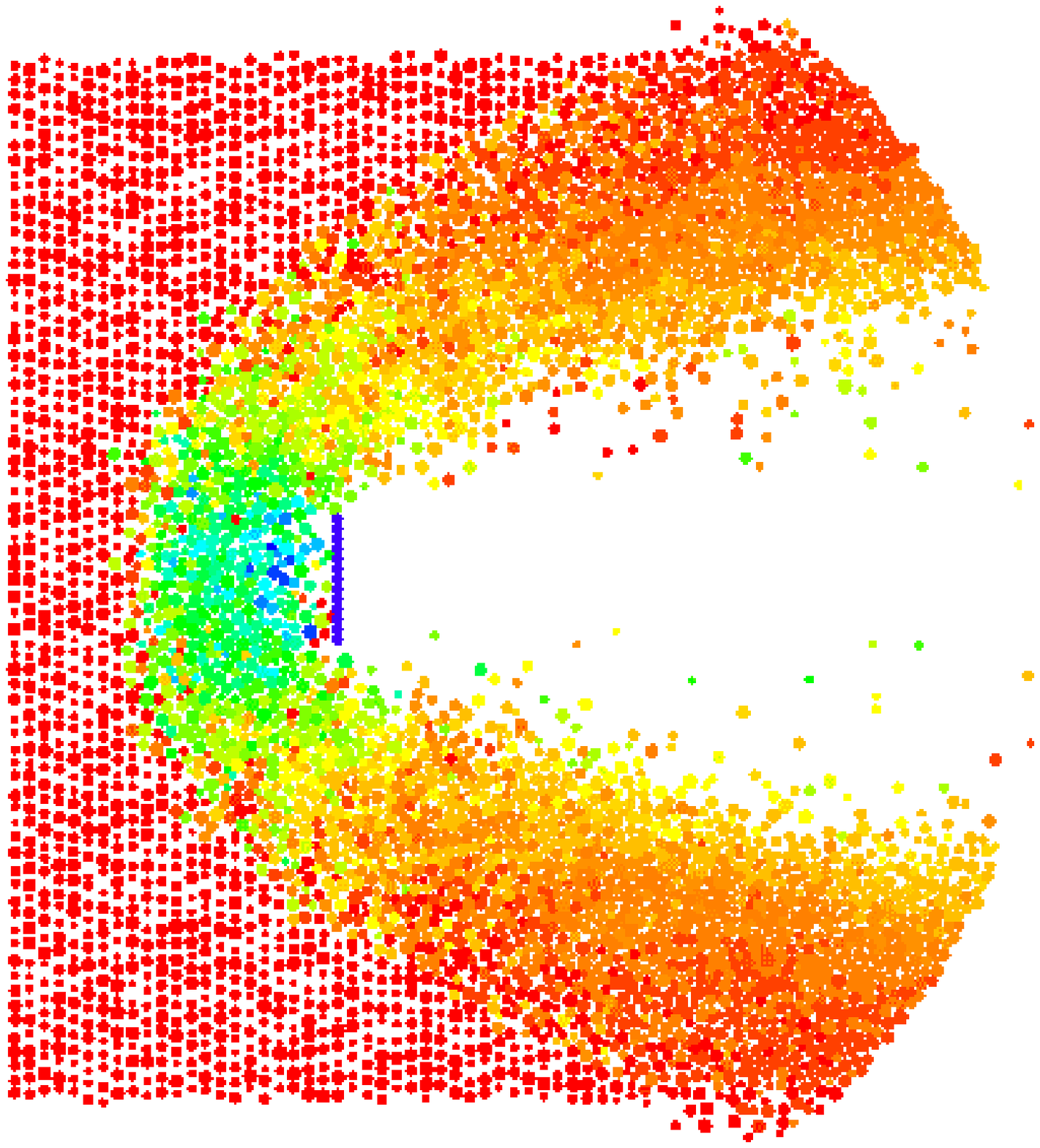,width=12cm,angle=0,clip=}} 
\vspace{0.3cm}
\caption{Snapshot of a simulation using a flat obstacle. Similar as in fig.~7
 one finds a corona of stream particles shielding the obstacle.}
\label{snapflat}
\end{figure}

\begin{figure}[htb]
\centerline{\psfig{figure=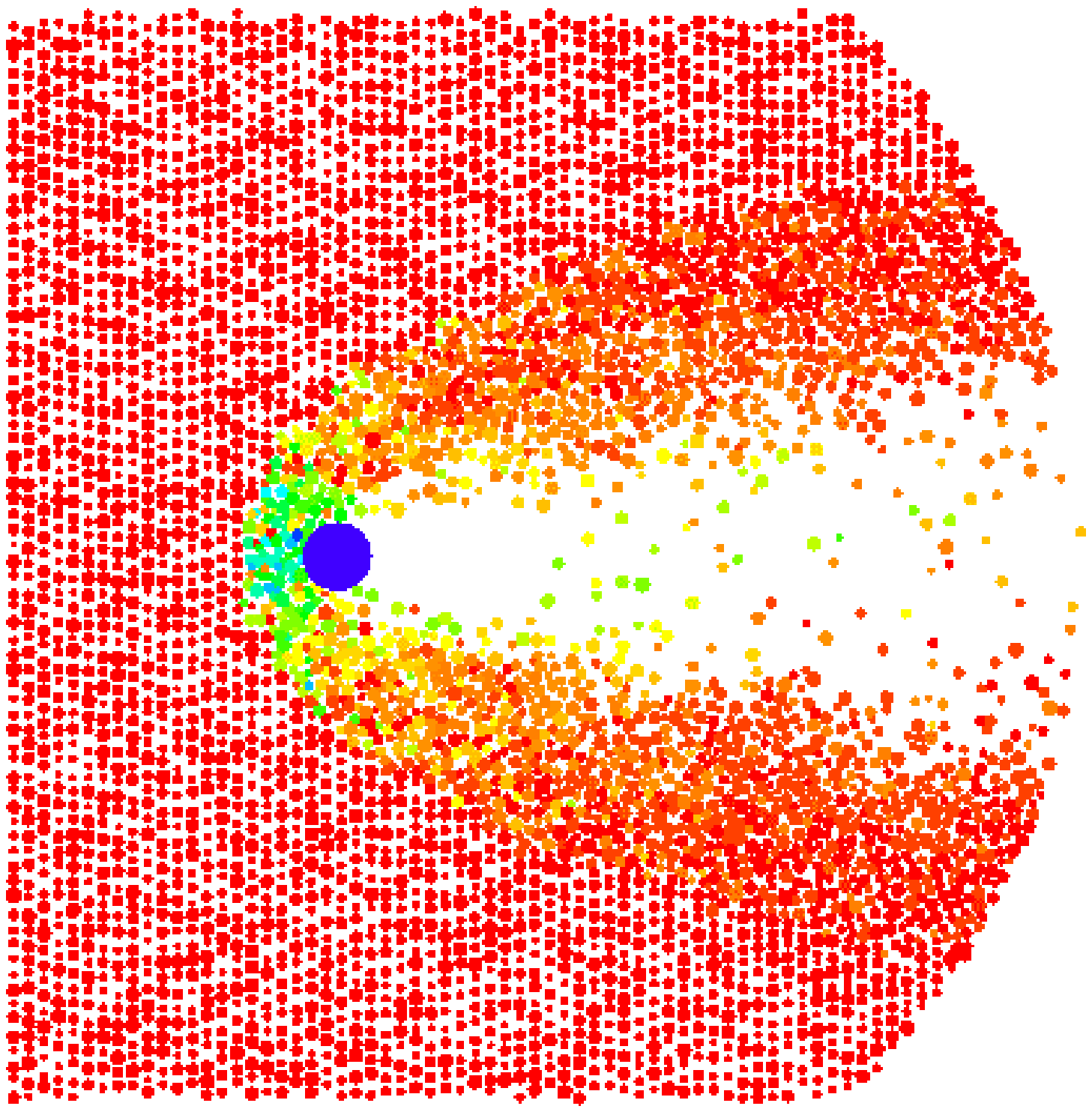,width=5.8cm,angle=0,clip=}\hspace*{0.2cm}\psfig{figure=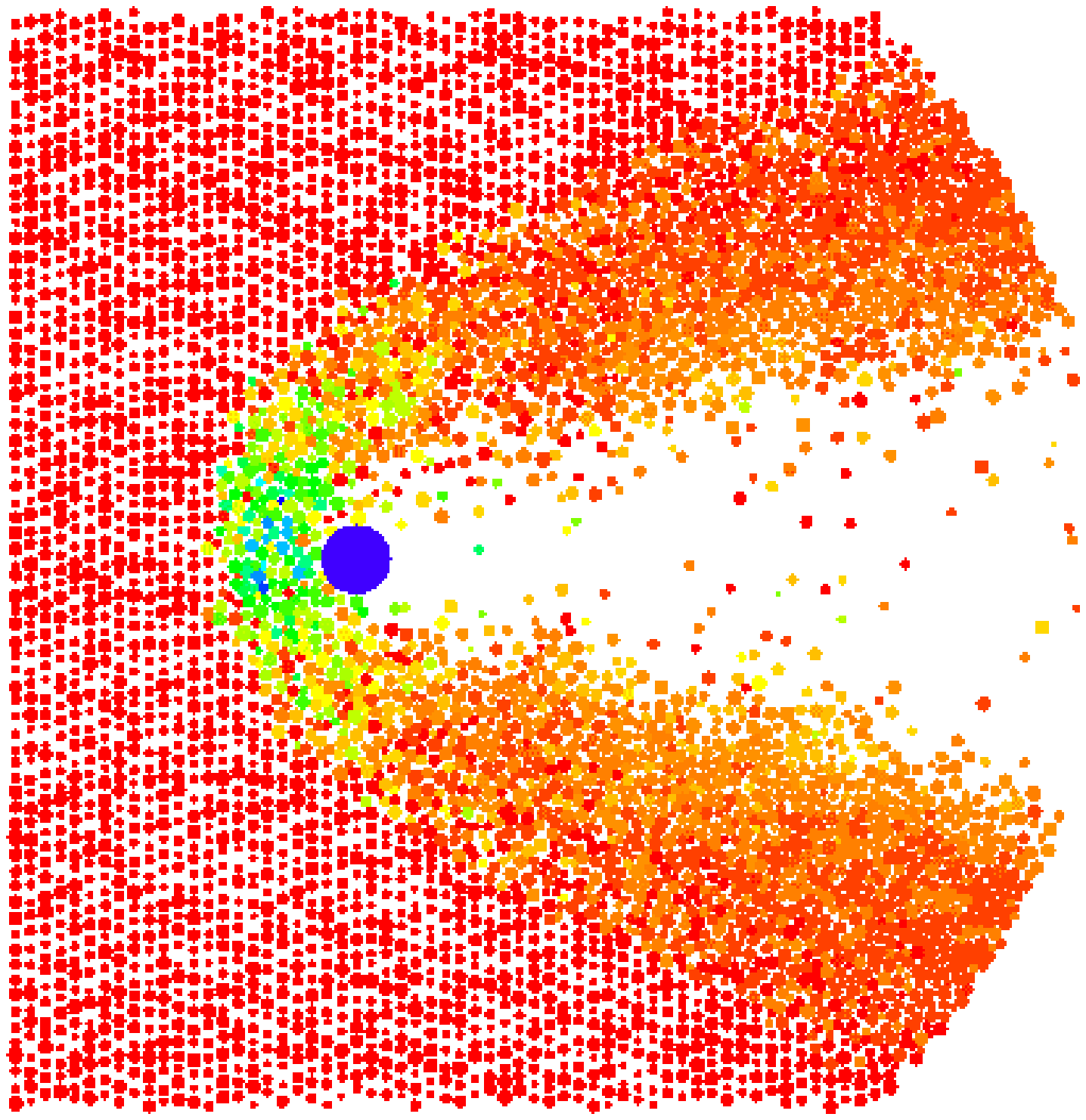,width=5.8cm,angle=0,clip=}\hspace*{0.2cm}\psfig{figure=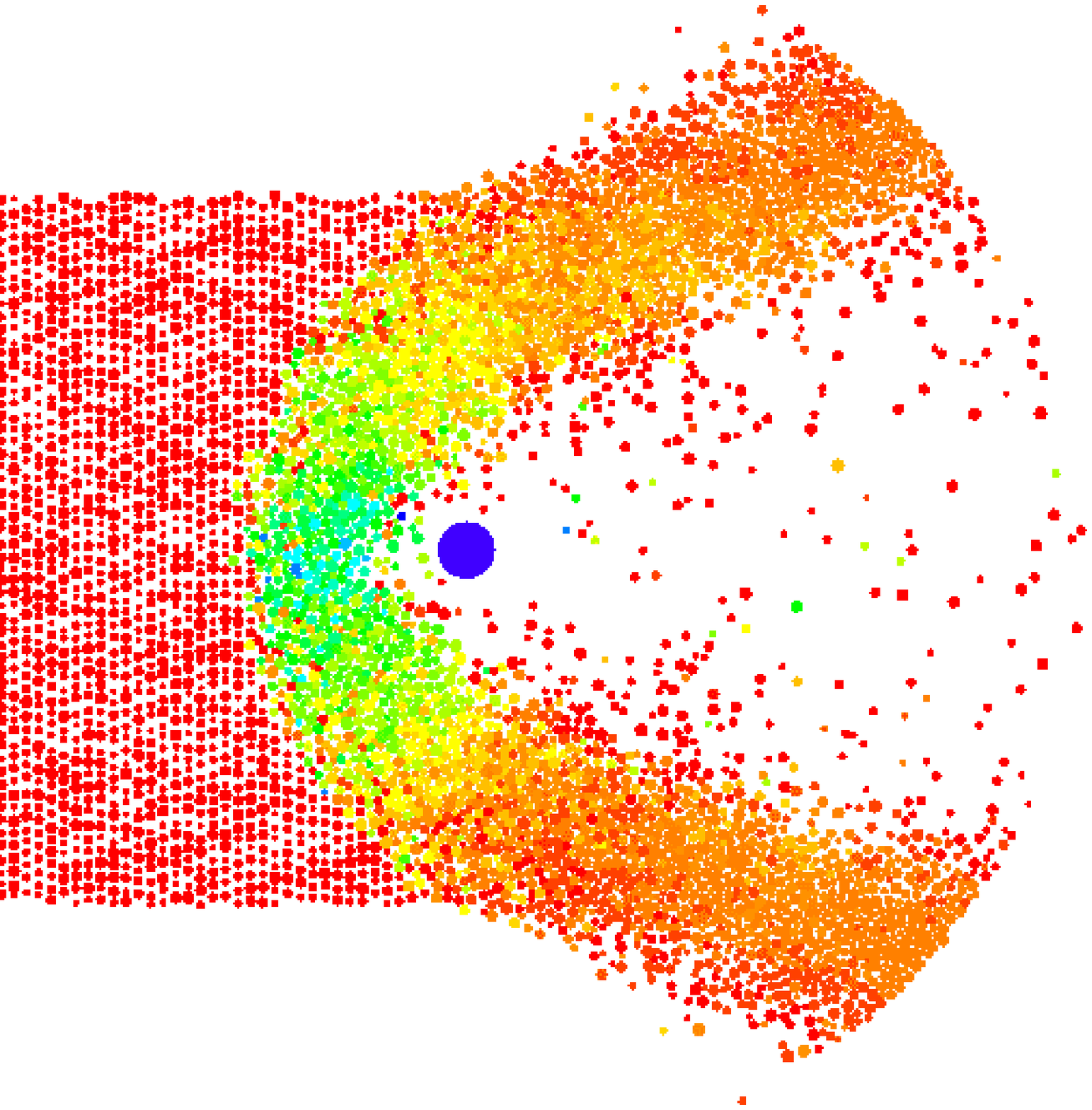,width=5.8cm,angle=0,clip=}} 
\vspace{0.3cm}
\caption{For lower impact velocity $v_{st}$ a gap in between the obstacle and the corona is observed. The gap appears at the same velocity when the force deviates from the quadratic law. Left: $v_{st}=2000\, \mbox{cm/sec}$, middle: $v_{st}=700\, \mbox{cm/sec}$, right: $v_{st}=200\, \mbox{cm/sec}$. Again the color codes for the absolute value of the particle velocities. The three figures are drawn with the same color scaling.}
\label{fig:gapsnaps}
\end{figure}

\begin{figure}[htb]
\centerline{\psfig{figure=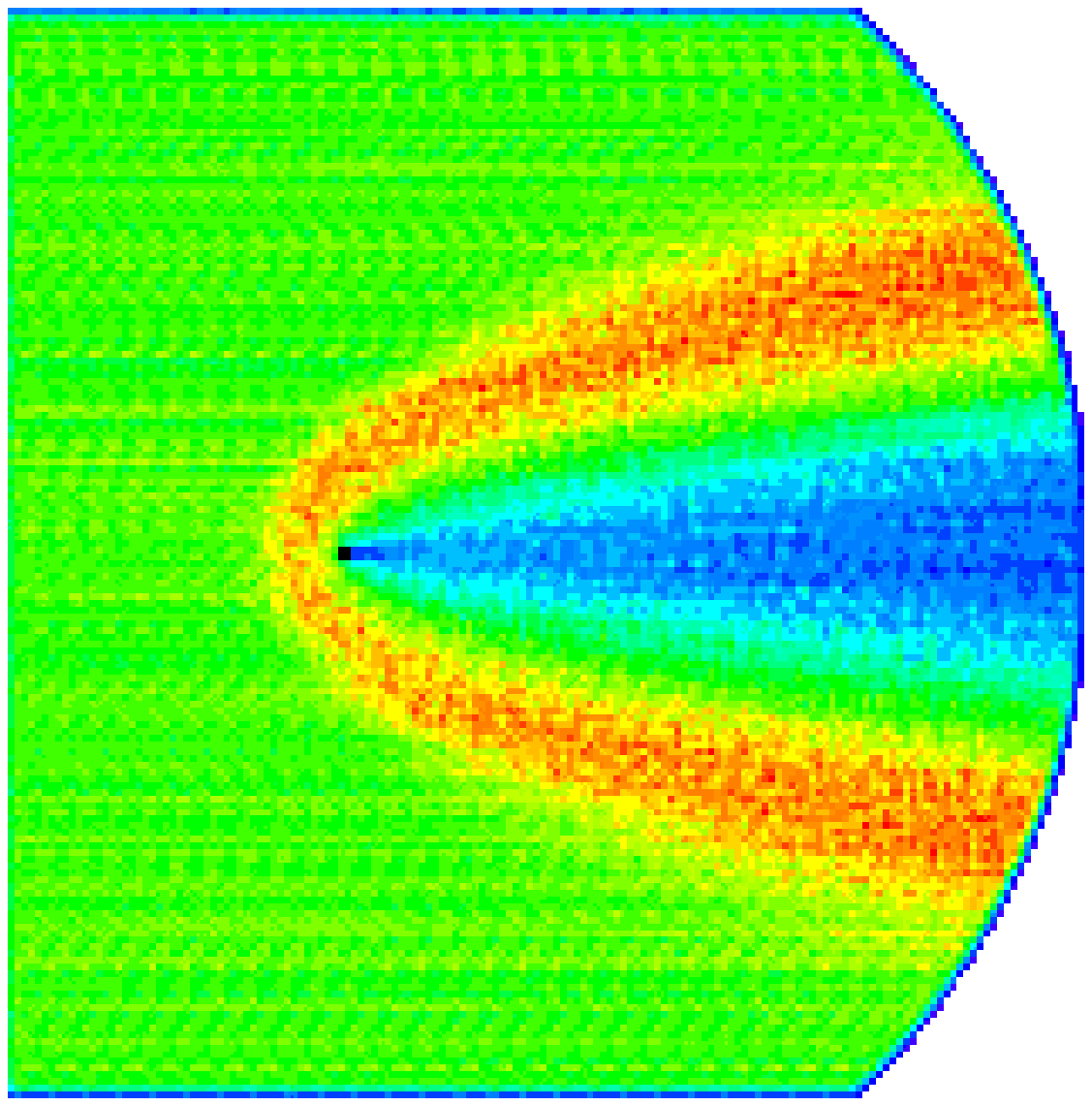,width=5.8cm,angle=0,clip=}\hspace*{0.2cm}\psfig{figure=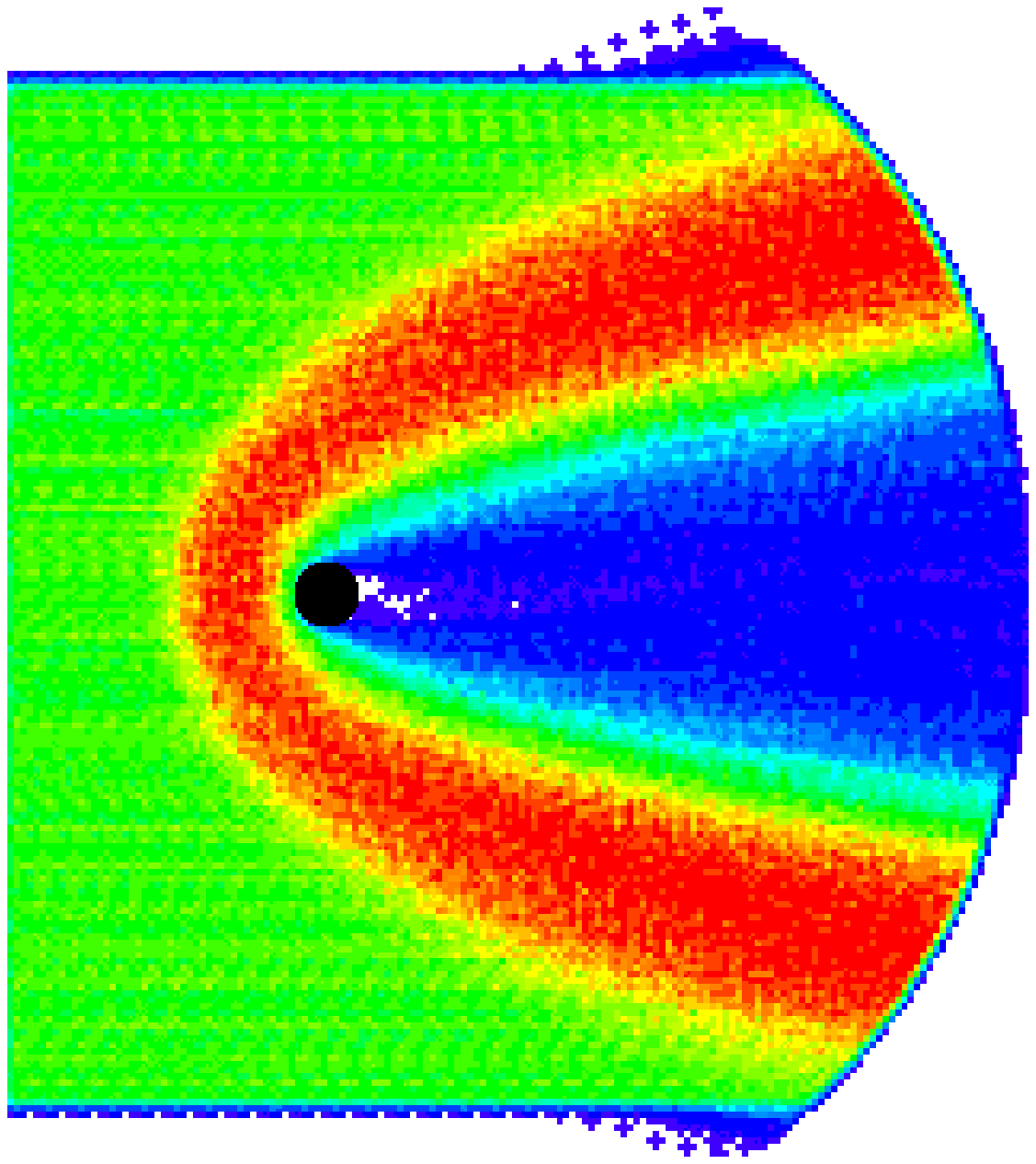,width=5.8cm,angle=0,clip=}\hspace*{0.2cm}\psfig{figure=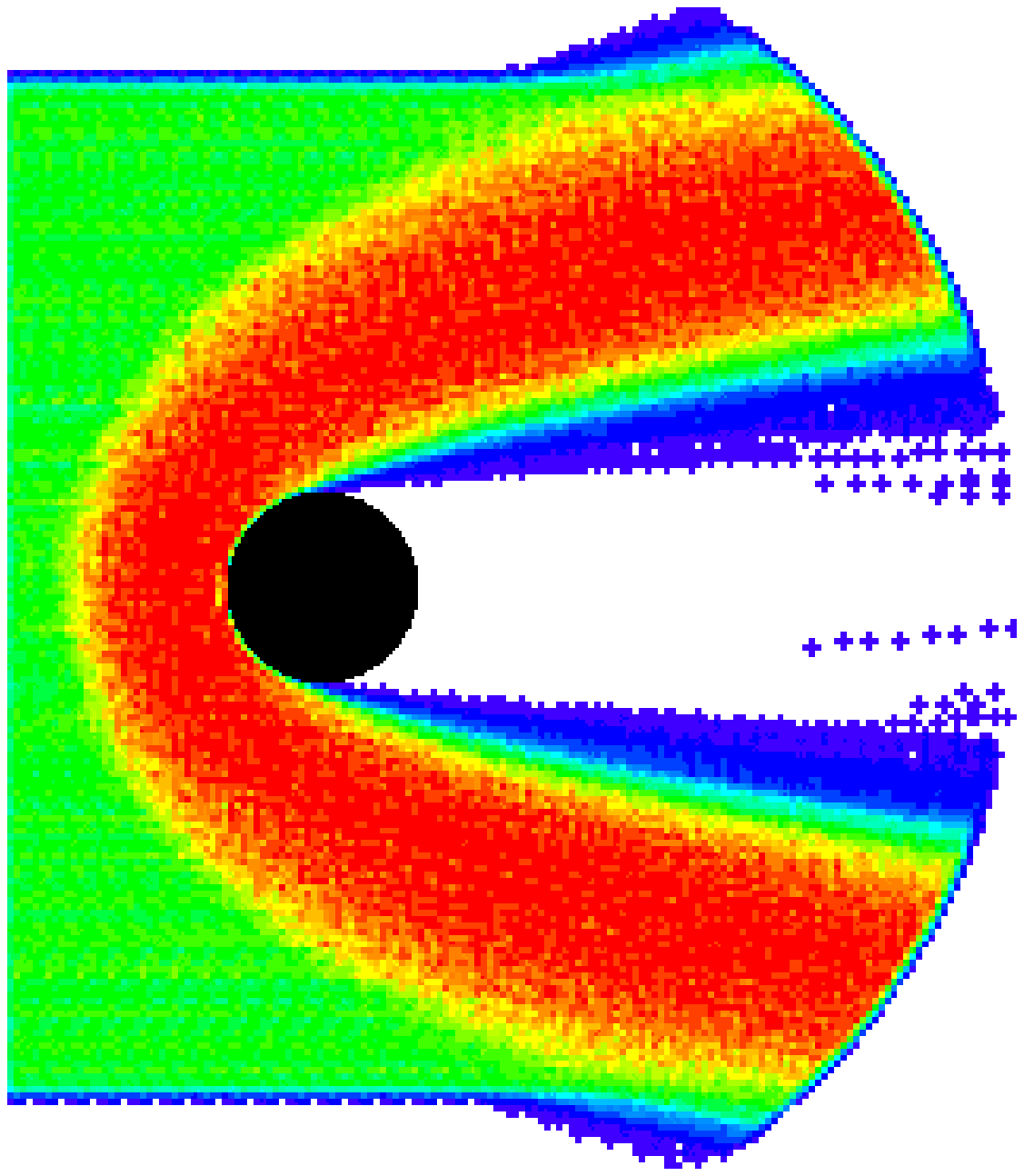,width=5.8cm,angle=0,clip=}} 
\vspace*{0.2cm}
\centerline{\psfig{figure=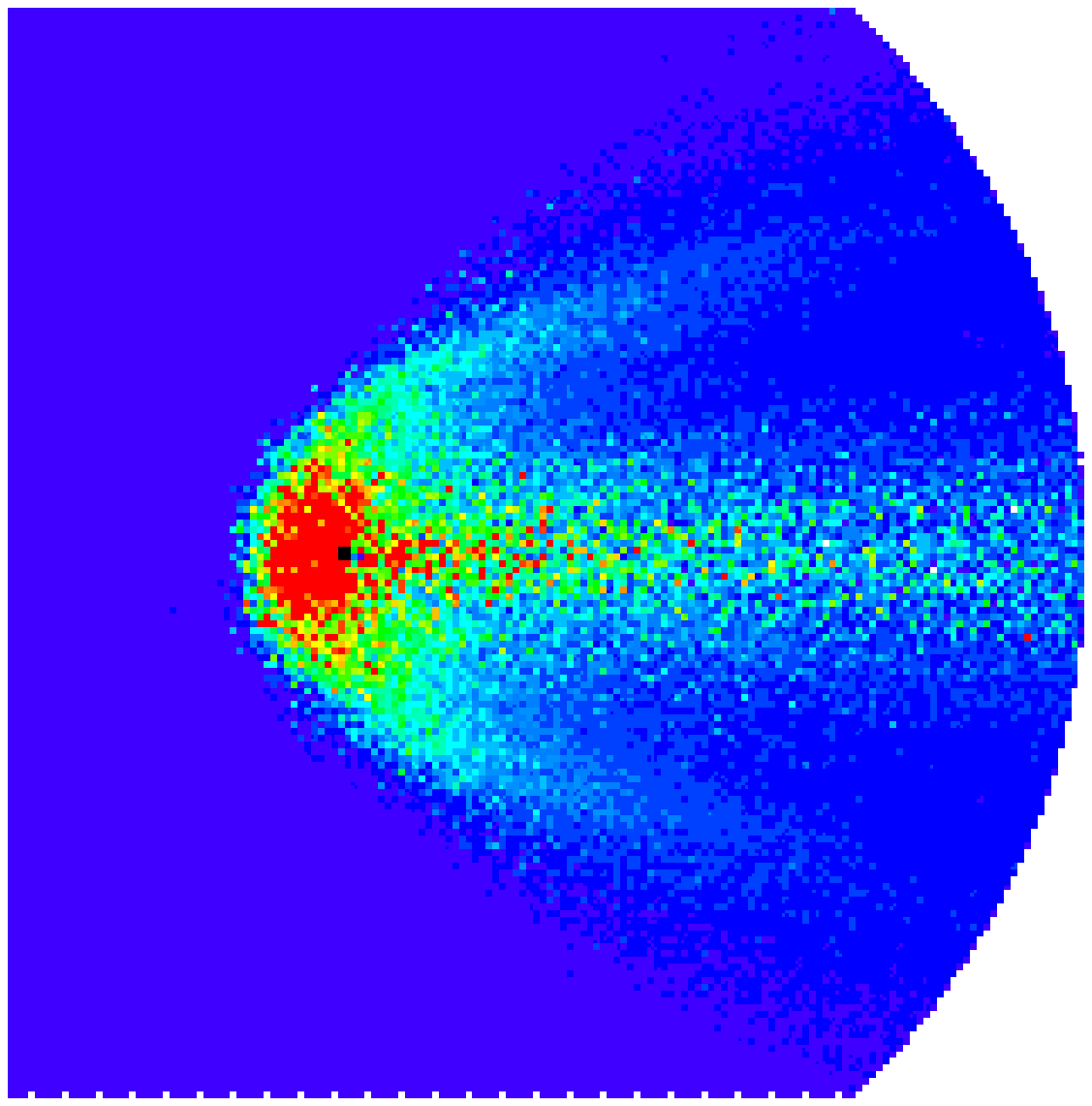,width=5.8cm,angle=0,clip=}\hspace*{0.2cm}\psfig{figure=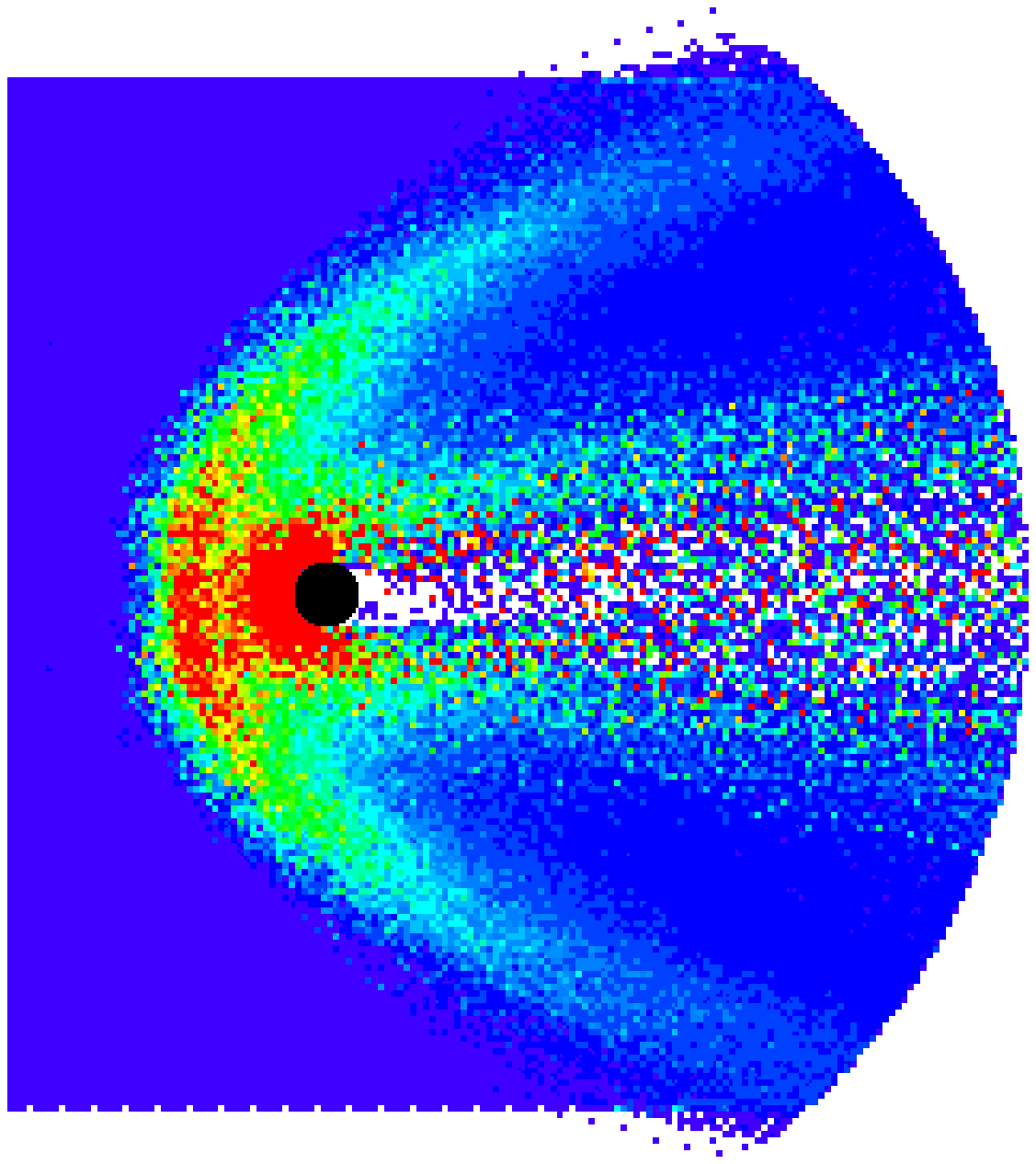,width=5.8cm,angle=0,clip=}\hspace*{0.2cm}\psfig{figure=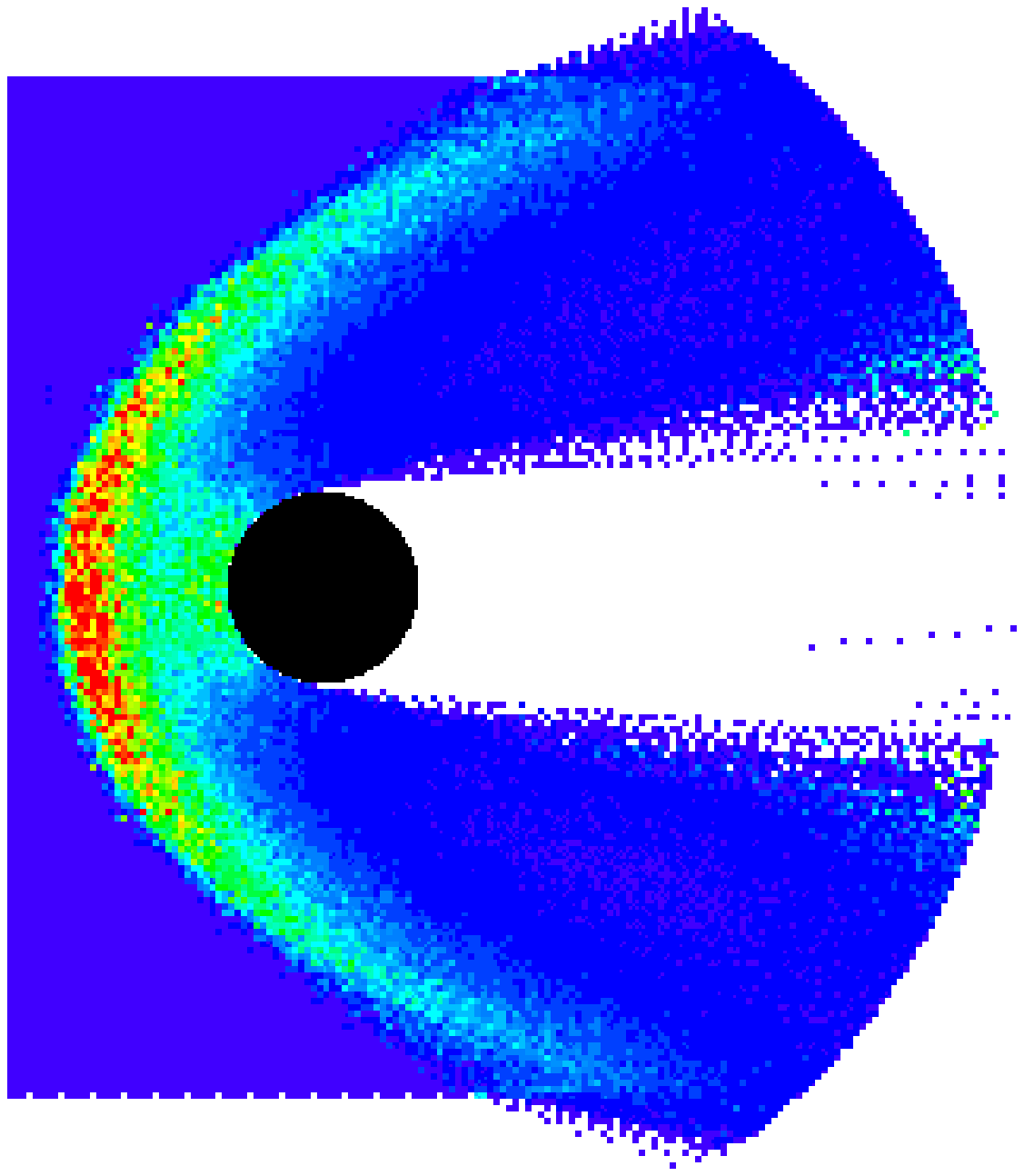,width=5.8cm,angle=0,clip=}} 
\vspace{0.3cm}
\caption{The fields of density (top row) and granular temperature (bottom row) for three different obstacle sizes. Discussion see text.}
\label{fig:fields}
\end{figure}

For the case of a sphere of radius $R$ in a viscous fluid one finds for very low stream velocity (low Reynolds number) experimentally the Stokes law $F = 6 \pi \eta\ R\,v_{st}$. This result can be derived analytically from the continuum mechanical Navier-Stokes equation for incompressible flow~\cite{StokesLaw}
\begin{equation}
\rho\left(\vec{v}_{st}\cdot \nabla \right) \vec{v}_{st} = -\nabla p + \eta \nabla^2 \vec{v}_{st}
\end{equation}
as result of the pressure in front and behind the obstacle. The conditions of low Reynolds number and high viscosity assure that streamlines of continuous media cling to each point of the obstacle, which is a condition for the derivation of the Stokes law. For the case of a granular material this condition does not hold, i.e. the pressure which acts from behind the obstacle is zero, or at least very close to zero since as visible from the snapshots figs.~\ref{fig}, \ref{snapflat} and~\ref{fig:gapsnaps} and from the density fields in fig.~\ref{fig:fields} there are almost no particles which collide with the obstacle from behind. According to these considerations we conclude that one cannot expect Stokes law even for very low stream velocity $v_{st}$ because the precondition of a homogeneous medium with high viscosity does not hold. 

The field data of the granular temperature and the density show that for large obstacle size the location of maximum temperature coincides with the boundary of the corona. For the case of smaller obstacles the highest density is found in a certain distance from the obstacle whereas the region close to the obstacle is filled by a diluted granular gas of high temperature. For fixed obstacle size this gap can be found below a certain stream velocity $v_{st}$. It grows rapidly for decreasing $v_{st}$.

Instead of using molecular dynamics technique for the simulation of the dynamics of the granular material which requires large computational effort, one could apply an ``event driven algorithm'' which works much faster than molecular dynamics. Using such algorithms one does not solve Newton's equation of motion for each single collision between the grains but one calculates the relative tangential and normal velocities after each collision as functions of the relative impact velocities (normal and tangential). Therefore one needs the normal and tangential viscoelastic restitution coefficients which are functions of the material constants and the impact velocities. For the approximation that the dilatation is independent of the impact velocities Pao and others analytically calculated the normal restitution coefficient~\cite{Pao,Goldsmith}. A more general approach for the derivation of the coefficients of restitution based on physical interaction forces between colliding spheres~\cite{BSHP,KuwabaraKono,MorgadoOppenheim:1997} has been given in \cite{SchwagerPoeschel:1996}. Effective algorithms for these type of event driven molecular dynamics simulations can be found e.g. in~\cite{Rapaport:1980,MarinRissoCordero:1993,ShidaAnzai:1992,LudingClementBlumenRajchenbachDuran:1994STUDIES,luding96e,Lubachevsky:1991}.

\acknowledgments 
We thank Joshua Colwell, Sergei E. Esipov, Hans J. Herrmann, Stefan Luding, Lutz Schimansky-Geier, Stefan Soko{\l}owski, and Frank Spahn for fruitful discussion.

\begin{center}
\begin{table}[htb]
\begin{tabular}{lll}
$\alpha = Y/(1-\nu^2)$  & $\alpha=10^{10}$ & $\mbox{g}\cdot \mbox{cm}^{-0.5} \cdot \mbox{sec}^{-2}$ \\
$\beta = A \cdot \alpha$ & $\beta=1000$ & $\mbox{g}\cdot \mbox{cm}^{-1}\cdot \mbox{sec}^{-1}$ \\
shear friction & $\gamma_S=1000$ & $\mbox{sec}^{-1}$ \\
Coulomb friction parameter & $\mu=0.5$ \\
material density & $\rho^{m} = 1$ & $\mbox{g}/\mbox{cm}^3$ \\
\end{tabular}
\caption{The parameters used in the simulations.}
\label{parameter}
\end{table}
  \end{center}

\newpage

\newpage

\begin{references}
\bibitem{Boltzmann} L.~Boltzmann, {\em Vorlesungen \"uber Gastheorie}, J. A. Barth (Leipzig, 1896-98); {\em Lecons sur la theorie des gaz}, Gauthier-Villars, (Paris, 1902-05).

\bibitem{BSHP} N.~V.~Brilliantov, F.~Spahn, J.-M.~Hertzsch, T.~P\"oschel, {\em A Model for Collisions in Granular Gases}, Phys. Rev. E {\bf 53}, 5382 (1996).

\bibitem{KuwabaraKono} G.~Kuwabara and K.~Kono, {\em Restitution coefficient in a collision between two spheres}, Jpn. J. Appl. Phys. {\bf 26}, 1230 (1987).

\bibitem{MorgadoOppenheim:1997}  W.~A.~M.~Morgado and I.~Oppenheim, {\em
Energy dissipation for quasielastic granular particle collisions}, Phys.~Rev.~E {\bf 55}, 1940 (1997).

\bibitem{HaffWerner} P.~K.~Haff and B.~T.~Werner, {\em Computer simulation of the mechanical sorting of grains},
Powder Technol. {\bf 48}, 239 (1986)

\bibitem{HW1} P.~A. Thompson and G.~S.~Grest, {\em Granular Flow: Friction and Dilatancy Transition}, Phys. Rev. Lett. {\bf 67}, 1751 (1991).

\bibitem{HW2} S.~Melin, {\em Wave Propagation in Granular Assemblies}, Phys. Rev. E {\bf 49}, 2353 (1994).

\bibitem{HW3} T.~P\"{o}schel and V.~Buchholtz, {\em Static friction phenomena in granular materials: Coulomb law versus particle geometry}, Phys. Rev. Lett. {\bf 71}, 3963 (1993).

\bibitem{HW4} G.~H.~Ristow and H.~J.~Herrmann,{\em Density Patterns in Granular Media}, Phys. Rev. E {\bf 50}, R5 (1994)

\bibitem{Gear} C.~W.~Gear, {\em Numerical Initial Value Problems in Ordinary Differential Equations}, Prentice-Hall, (Englewwods Cliffs, 1971).

\bibitem{Allen} M.~P.~Allen and D.~J.~Tildesley, {\em Computer Simulations of Liquids}, Clarendon Press (Oxford, 1987).

\bibitem{EPR} S. E. Esipov, T. P\"oschel, and D. Rosenkranz, {\em Granular gas in the presence of gravity}, (preprint, 1997).

\bibitem{Haff:1983} P.~K.~Haff, {\em Grain flow as a fluid-mechanical phenomenon}, J.~Fluid~Mech. {\bf 134}, 401 (1983). 

\bibitem{GoldhirschZanetti:1993} I. Goldhirsch and G. Zanetti, {\em Clustering Instability in Dissipative Gases}, Phys. Rev. Lett. {\bf 70}, 1619 (1993).

\bibitem{GoldhirschSela} I. Goldhirsch and N. Sela, {\em Hydrodynamic equations for rapid shear flow}, J. Fluid Mech. (in press).

\bibitem{EsipovPoeschel:1995} S. E. Esipov and T. P\"oschel, {\em The Granular Phase Diagram}, J. Stat. Phys {\bf 86}, 1385 (1997).

\bibitem{LunSavage:1987} C. K. K. Lun and S. B. Savage, {\em A simple kinetic theory for granular flow of rough, inelastic, spherical particles}, J. Appl. Math. {\bf 54}, 47 (1987).

\bibitem{SEP} C. Salue\~{n}a, S. E. Esipov, and T. P\"oschel, {\em Hydrodynamic fluctuations and averaging problems in dense granular flows}, In: R. P. Behringer and J. T. Jenkins, {\em Powders and Grains'97}, p. 341, Balkema (Rotterdam, 1997).

\bibitem{SPE} C. Salue\~na, T. P\"oschel, and S. E. Esipov, {\em Dissipative properties of vibrated granular materials}, (preprint, 1997).

\bibitem{TanGoldhirsch} M-L. Tan and I. Goldhirsch, {\em Subsonic and supersonic regions in rapid granular flows}, (preprint)

\bibitem{EsipovPC} S. E. Esipov, {\em personal communication}.

\bibitem{GoldshteinShapiroGutfinger3:1996} A.~Goldshtein, M.~Shapiro, and C.~Gitfinger, {\em Mechanics of collisional motion of granular materials. Part 3: Self-similar shock wave propagation}, J. Fluid Mech. {\bf 316}, 29 (1996).

\bibitem{BEP} V.~Buchholtz, S.~E.~Esipov, and T.~P\"oschel, {\em in preparation}

\bibitem{StokesLaw} L.~D.~Landau and E.~M.~Lifshitz, {\em Fluid Mechanics}, Pergamon Press (London,1959).

\bibitem{Pao} Y.-H.~Pao, {\em Extension of the Hertz theory of impact to the viscoelastic case}, J.Appl.Phys. {\bf 26}, 1083 (1955).

\bibitem{Goldsmith} W.~Goldsmith, {\em Impact: The Theory and Physical Behaviour of Colliding Solids}, E.~Arnold (London,1960).

\bibitem{SchwagerPoeschel:1996} T.~Schwager and T.~P\"oschel, {\em Coefficient of resitution of viscous particles and cooling rate of granular gases}, Phys. Rev. E, in press (1997).

\bibitem{Rapaport:1980} D. C. Rapaport, {\em The event scheduling problem in molecular dynamic simulations}, J. Comp. Phys. {\bf 34}, 184 (1980).

\bibitem{MarinRissoCordero:1993} M.~Mar\'{\i}n, D.~Risso, and P.~Cordero, {\em Efficient algorithms for many body hard particle molecular dynamics}, J. Comp. Phys. {\bf 109}, 306.

\bibitem{ShidaAnzai:1992} K.~Shida and Y.~Yuichiro, {\em Reduction of the event-list for molecular dynamic simulation}, Comp. Phys. Comm. {\bf 69}, 318 (1992).

\bibitem{LudingClementBlumenRajchenbachDuran:1994STUDIES} S.~Luding, E.~Cl\'ement, A.~Blumen, J.~Rajchenbach, and J.~Duran, {\em Studies of columns of beads under external vibration},  Phys. Rev. E {\bf 49}, 1634 (1994).

\bibitem{luding96e} S.~Luding, E.~Cl\'ement, J.~Rajchenbach, and J.~Duran, {\em Simulations of pattern formation in vibrated granular media}, Europhys. Lett. {\bf 36}, 247 (1996).

\bibitem{Lubachevsky:1991} B.~D.~Lubachevsky, {\em How to Simulate Billiards and Similar Systems}, J. Comp. Phys. {\bf 94}, 255 (1991).


\end{references}
\end{document}